\documentclass[aps,pra,10pt,letterpaper,twocolumn,superscriptaddress]{revtex4-1}%
\usepackage{amsfonts}
\usepackage{amsmath}
\usepackage{amssymb}
\usepackage{graphicx}
\usepackage{natbib}%
\providecommand{\U}[1]{\protect\rule{.1in}{.1in}}

\begin{document}

\title{Hyperspherical approach to the three-bosons problem in 2D with a magnetic field.}
\author{Seth T. Rittenhouse}
\affiliation{Department of Physics, The United States Naval Academy, Annapolis, MD 21402, USA}
\affiliation{Department of Physics and Astronomy, Western Washington University, Bellingham, WA 98225, USA}
\author{Andrew Wray}
\affiliation{Department of Physics and Astronomy, Western Washington University, Bellingham, WA 98225, USA}
\author{B. L. Johnson}
\affiliation{Department of Physics and Astronomy, Western Washington University, Bellingham, WA 98225, USA}

\begin{abstract}
We examine a system of three-bosons confined to two dimensions in the presence of a perpendicular magnetic field within the framework of the adiabatic hyperspherical method.  For the case of zero-range, regularized pseudo-potential interactions, we find that the system is nearly separable in hyperspherical coordinates and that, away from a set of narrow avoided crossings, the full energy eigenspectrum as a function of the 2D s-wave scattering length is well described by ignoring coupling between adiabatic hyperradial potentials.  In the case of weak attractive or repulsive interactions, we find the lowest three-body energy states exhibit even/odd parity oscillations as a function of total internal 2D angular momentum and that for weak repulsive interactions, the universal lowest energy interacting state has an internal angular momentum of $M=3$.  With the inclusion of repulsive higher angular momentum we surmise that the origin of a set of ``magic number'' states (states with anomalously low energy) might emerge as the result of a combination of even/odd parity oscillations and the pattern of degeneracy in the non-interacting lowest Landau level states.

\end{abstract}
\date{\today}

\maketitle

\affiliation{Department of Physics and Astronomy, Western Washington University,
Bellingham, WA 98225, USA}

\affiliation{Department of Physics and Astronomy, Western Washington University,
Bellingham, WA 98225, USA}

\affiliation{Department of Physics and Astronomy, Western Washington University,
Bellingham, WA 98225, USA}

\section{Introduction}

The discovery of the fractional quantum Hall effect (FQHE)\cite{fqhe1} provided an important
impetus to investigate the physics of low-dimensional systems wherein particle-
particle interactions play a fundamental role in the resulting dynamics. Since the
original work of Laughlin\cite{laughlin1}, many theoretical approaches have been employed both to
explain the observed collective behavior  in the FQHE regime, and to expand the
tools utilized for, and make connections between, a variety of strongly-correlated
many-body problems. In particular, model wavefunctions (building upon the
original variational ground state proposed by Laughlin) have been employed to
study the ground state and elementary excitations\cite{BoYang,Wu1}, as well as composite-particle
models\cite{Wilczek1,Wilczek2,Jain1}, exact diagonalization for few-body systems\cite{girvin1983interacting,
Lai1,stone1,johnson1} and special conformal field
theories\cite{Read1,Gurari1,Bond1}. The problem is extremely rich; although much early work focused on
interacting fermions, the generalization of the problem to other systems has been
ongoing.

The richness associated with the fundamental dynamics of the FQHE regime (i.e. a
system of two-dimensional (2D) charged particles under a large transverse DC
magnetic field) may be traced to the dynamic bound states resulting from the
interplay of the inter-particle interactions and the external magnetic field. In 2D, any
repulsive interaction will serve to separate the constituent particles to lower the
energy, and the resulting separating motion will couple to the magnetic field, turning
the particle trajectories back on themselves. The resulting dynamic many-particle bound
state (in the absence of additional confining potentials),
is a rich problem,
even classically, wherein a primary feature is that the bound system is dominated by
rotation, and thus the system angular momentum is a key feature \cite{johnson2,Fine1,johnson3}. A useful tool for
connecting the classical problem to the quantum many-body problem is the study of
the two-body FQH problem\cite{Ge1}; here the problem is separable, and the precise role of
the relative angular momentum is evident.

In the absence of inter-particle interactions, the spectrum of a 2-D charged gas
under transverse field is comprised of a set of highly-degenerate Landau levels\cite{johnson4}. In
the FQHE regime, the dynamic bound states described above are responsible for
creating gaps in the energy spectrum {\it within} the Landau levels, and it is the
origin and structure of these gaps, as well as the elementary excitations associated
with them, that has garnered significant attention.

It was originally pointed out by Laughlin\cite{Laughlin2}, and later studied in detail for few-body
systems, that in the ground state of  the interacting system in the lowest Landau
level the angular momentum for small clusters goes up in integer multiples of the
particle number \cite{girvin1983interacting}. In addition, exact diagonalization studies of few-body clusters
interacting via Coulomb potentials exhibit `magic numbers' in ground state energy
of the system as a function of the total angular momentum\cite{girvin1983interacting,johnson1}; at specific values of total angular
momenta, local minima appear in the ground state energy, and at specific values of
the `magic' angular momenta, a corresponding maxima in the the excitation gap
from the ground state appears\cite{girvin1983interacting}. The magic angular momenta for which the gaps
appear are precisely those given by the trial ground-state wavefunction, and may be
interpreted via the observation\cite{johnson1} that in the FHQE ground state, the relative angular
momentum for each pair of particles is the {\it same}, and the total angular
momentum is therefore found via $M=\sum m_{ij}=[N(N-1)/2]q$, the usual result
for the thermodynamic limit of large systems. Here $M$ is the total angular
momentum, $m_{ij}$ is the relative angular momentum quantum number for
particles $i$ and $j$, $N$ is the total number of particles, and $q$ is an odd integer
(Fermi statistics). The origin of this remarkable result, and its dependence upon the
form of the interactions, is an interesting question.

A further interesting question is whether or not the features of the FQHE regime
generalize in a straightforward manner. In particular, can the structure of the
vector potential and the particle interactions be projected in order to illuminate
physical reasons for some of the fundamental structures? Recently, separation of
the Schr\"{o}dinger  equation in hyperspherical coordinates in the FQHE regime
demonstrated that the appearance of the FQHE may be associated with patterns in
the degeneracy of the states\cite{daily2015hyperspherical}, a conjecture also made earlier by Stone, et. al.\cite{stone1}. 

In this paper, we study the dynamics of
few-body 2D bosonic systems interacting via regularized pseudo-potential contact interactions under transverse magnetic field within the adiabatic hyperspherical method. The hyperspherical framework gives a convenient picture for studying the nature of the role of interactions and the degeneracy of
the ground state, as well as the resulting Landau-level structure.  When interactions are included, we find that, away from a set of narrow avoided crossings in the eigenspectrum, the three-boson problem is nearly separable in hyperspherical coordinates, in agreement with previous results for few-electrons interacting via the Coulomb interaction \cite{daily2015hyperspherical}.  Weak attractive or repulsive s-wave interactions also produce ground state energies for fixed angular momentum that exhibit even/odd parity oscillations. On the repulsive side, we surmise that if repulsive higher angular momentum interactions are included, this parity oscillation combined with the degeneracy pattern of the lowest Landau level might be the source of the `magic number' behavior predicted in previous studies.

This paper is organized as follows.  In Section II we describe our theoretical approach including developing the Hamiltonian for three-bosons in 2D with a perpendicular magnetic field, separating out the center of mass motion, and transforming into hyperspherical coordinates.  We also describe the Landau level structure that results for the noninteracting case and develop a transcendental equation, the roots of which produce the adiabatic hyperradial potentials.  In Section III we examine the resulting adiabatic potentials and discuss the limiting cases of large and small s-wave two-body scattering length.  In Section IV, the full eigenspectrum of the three-boson problem is presented as well as an analysis of the ground state energy in the weakly repulsive and attractive limits.  In Section V, the role of degeneracy is discussed as well as the part that degeneracy might play in the emergence of magic numbers in the three-boson system.  Finally, in Section VI we summarize the results presented in the paper.

\section{Theoretical methods}

The few-body Hamiltonian that we are concerned with is that of three
identical bosons confined to two dimensions in a vector potential appropriate to a constant effective magnetic field perpendicular to the plane of motion:%
\begin{align}
H  &  =\sum_{i=1}^{3}h_{i}+\sum_{i<j}V\left(  r_{ij}\right)
.\label{Eq:TotHam}\\
h_{i}  &  =\dfrac{1}{2m}\left(  -i\hbar\nabla_{i}+\alpha\vec{A}_{i}\right)
^{2}\nonumber
\end{align}
where $h_{i}$ is the single particle Hamiltonian for a particle moving in the
vector potential $\vec{A}$, and $r_{ij}=\left\vert \vec{r}_{i}-\vec{r}_{j}\right\vert $ is
the inter-particle separation distance between particles $i$ and $j$. Here
$\vec{A}_{i}$ is the vector potential experienced by the $i$th particle, and
$\alpha$ is an overall scaling factor. If the particles in question are
charged particles, the scale factor would be simply given by $\alpha=q/c$ in
gaussian units. In Eq. \ref{Eq:TotHam}, $V\left(  r\right)  $ is a pairwise
isotropic interaction between two bosons that will be described more fully
below. Since $\vec{A}$ creates an effective constant magnetic field, here we
choose to describe this field in the symmetric gauge:
\begin{equation}
\vec{A}=\dfrac{B_{0}}{2}\left(  \hat{x}y-\hat{y}x\right)  . \label{Eq:Vecpot}%
\end{equation}
Note that we have chosen the magnetic field to be pointing in the $-z$
direction. Inserting the vector potential into Eq. \ref{Eq:TotHam} gives a
total Hamiltonian in an illuminating form:%
\begin{equation}
H=\sum_{i=1}^{3}\left(  \dfrac{-\hbar^{2}}{2m}\nabla_{i}^{2}+\dfrac{1}%
{8}m\omega_{c}r_{i}^{2}\right)  -\dfrac{\omega_{c}}{2}L_{z,Tot}+\sum
_{i<j}V\left(  r_{ij}\right)  . \label{Eq:TotHamexpand}%
\end{equation}
where $L_{z,Tot}=\sum_{i}\ell_{z,i}$ is the total angular momentum of the
system. Here, we have written the Hamiltonian in terms of the
cyclotron frequency $\omega_{c}=\alpha B/m$. The cyclotron frequency also
yields a length scale $l_{c}=\sqrt{\hbar/m\omega_{c}}=\sqrt{\hbar/\alpha B}$
called the magnetic length. The utility of the symmetric gauge is now obvious:
the effect of the magnetic field is simply that of an isotropic trap in the
system along with an overall shift downward determined by the total angular
momentum of the system.

To separate out the center of mass, we transform into a set of mass-scaled
Jacobi coordinates \cite{delves1959tertiary,delves1960tertiary}:%
\begin{align}
\vec{\rho}_{1}^{\left(  k\right)  }  &  =\sqrt{\dfrac{\mu_{i,j}}{\mu}}\left(
\vec{r}_{i}-\vec{r}_{j}\right)  , \label{Eq:jaccoords}\\
\vec{\rho}_{2}^{\left(  k\right)  }  &  =\sqrt{\dfrac{\mu_{ij,k}}{\mu}}\left(
\dfrac{\vec{r}_{i}+\vec{r}_{j}}{2}-\vec{r}_{k}\right)  , \nonumber \\
\vec{R}_{CM}  &  =\dfrac{\vec{r}_{1}+\vec{r}_{2}+\vec{r}_{3}}{3} \nonumber%
\end{align}
where $\mu_{1,2}=m/2$ is the 2-body reduced mass, $\mu_{ij,k}=2m/3$ is the
reduced mass of a two body system with third particle and $\mu$ is the
three-body reduced mass which we choose to be $\mu=m/\sqrt{3}$. Here the
superscript $\left(  k\right)  $ indicates which Jacobi coordinates have been
chosen using the \textquotedblleft odd-man out\textquotedblright\ notation where ${i,j,k}$ is a cyclic permutation of the particle numbers e.g. if $k=3$ then $i=1$ and $j=2$. After
the transformation the total Hamiltonian can be written as%
\begin{align}
H=  &  H_{int}+H_{CM},\label{Eq:Hcmsep}\\
H_{CM}=  &  -\dfrac{\hbar^{2}}{2M_{Tot}}\nabla_{CM}^{2}+\dfrac{1}{8}M_{Tot}\omega_{c}%
^{2}R_{CM}^{2}-\dfrac{\omega_{c}}{2}L_{CM,z},\nonumber\\
H_{int}=  &  -\dfrac{\hbar^{2}}{2\mu}\left(  \nabla_{1}^{2}+\nabla_{2}%
^{2}\right)  -\dfrac{\omega_{c}}{2}L_{int,z}\label{Eq:Hint1}\\
&  +\dfrac{1}{8}\mu\omega_{c}^{2}\left(  \rho_{1}^{2}+\rho_{2}^{2}\right)
+\sum_{i<j}V\left(  r_{ij}\right)  .\nonumber
\end{align}
Here $L_{CM,z}$ is the angular momentum operator of the center of mass, $L_{int,z}$
is the internal angular momentum operator, and $M_{Tot}=3m$ is the total mass of the
three-body system. In Eq. \ref{Eq:Hint1}, $\nabla_{i}$ refers to a derivative
with respect to the $i$th Jacobi coordinate. It is important to point out that
the internal Hamiltonian has the same form, independent of which Jacobi
coordinate system has been chosen from Eq.~\ref{Eq:jaccoords} and thus the superscript $(k)$ has been suppressed.

Since the center of mass motion is completely separated, we can proceed to
examine the internal Hamiltonian, $H_{int}$. Because the interactions are
isotropic, the total internal 2D angular momentum of the system is a good
quantum number. If we restrict the system to only states with total internal
angular momentum $M$, the Schr\"{o}dinger equation that results from Eq.
\ref{Eq:Hint1} is that of three particles confined to an isotropic oscillator
with oscillator frequency $\omega_{c}/2$, i.e.%
\begin{align}
E\Psi_{M}\left(  \vec{\rho}_{1}^{\left(  k\right)  },\vec{\rho}_{2}^{\left(
k\right)  }\right)  =  &  \left[  \sum_{i}\left(  \dfrac{-\hbar^{2}}{2\mu
}\nabla_{i}^{2}+\dfrac{1}{8}\mu\omega_{c}^{2}\rho_{i}^{2}\right)  \right.
\label{Eq:SE1}\\
&  +\left.  \sum_{i<j}V\left(  r_{ij}\right)  -\dfrac{\omega_{c}}{2}M\right]
\Psi_{M}\left(  \vec{\rho}_{1}^{\left(  k\right)  },\vec{\rho}_{2}^{\left(
k\right)  }\right) \nonumber
\end{align}
where the first sum on the right hand side runs over the Jacobi vectors. Here
$\Psi_{M}\left(  \vec{\rho}_{1}^{\left(  k\right)  },\vec{\rho}_{2}^{\left(
k\right)  }\right)  $ is a three-body eigenfunction with total internal
angular momentum $M$.

\subsection{Hyperspherical coordinates}

To solve Eq. \ref{Eq:SE1} we employ hyperspherical coordinates and the
adiabatic hyperspherical representation, wherein the 4-dimensional
Schr\"{o}dinger equation is expressed in terms of the hyperradius
$R=\sqrt{\rho_{1}^{2}+\rho_{2}^{2}}$ and a set of three hyperangles $\left\{
\alpha,\phi_{1},\phi_{2}\right\}  ,$ collectively denoted by $\Omega,$ where
$\phi_{1}$ and $\phi_{2}$ are the standard polar angles for Jacobi vectors
$\vec{\rho}_{1}$ and $\vec{\rho}_{2}$ respectively, and $\alpha$ is an angle
that correlates the lengths of the two Jacobi vectors, i.e.%
\begin{align*}
\rho_{1}  &  =R\sin\alpha,\\
\rho_{2}  &  =R\cos\alpha.
\end{align*}
In hyperspherical coordinates, the internal Hamiltonian of Eq. \ref{Eq:Hint1}
can be written as%
\begin{align}
H_{int}=  &  \dfrac{-\hbar^{2}}{2\mu}\left(  \dfrac{1}{R^{2}}\dfrac{\partial
}{\partial R}R^{3}\dfrac{\partial}{\partial R}-\dfrac{\Lambda^{2}}{R^{2}%
}\right)  +\dfrac{i\hbar\omega_{c}}{2}\left(  \dfrac{\partial}{\partial
\phi_{1}}+\dfrac{\partial}{\partial\phi_{2}}\right) \label{Eq:HintHS}\\
&  +\dfrac{1}{8}\mu\omega_{c}^{2}R^{2}+\sum_{i<j}V\left(  r_{ij}\right)
,\nonumber\\
\Lambda^{2}=  &  \dfrac{-1}{\sqrt{\sin\alpha\cos\alpha}}\dfrac{\partial^{2}%
}{\partial\alpha^{2}}\sqrt{\sin\alpha\cos\alpha}\nonumber\\
&  -\dfrac{1}{\sin^{2}\alpha}\left(  \dfrac{\partial^{2}}{\partial\phi_{1}%
^{2}}-\dfrac{1}{4}\right)  -\dfrac{1}{\cos^{2}\alpha}\left(  \dfrac
{\partial^{2}}{\partial\phi_{2}^{2}}-\dfrac{1}{4}\right)  .\nonumber
\end{align}
Here $\Lambda^{2}$ is the grand angular momentum operator, the properties and
description of which can be found in a number of references (see Refs.
\cite{smirnov1977method,avery1989hyperspherical,avery1993selected} for example).

\subsection{Landau levels}

Before we solve the fully interacting system, it is instructive
to consider the structure of the solutions to the non-interacting system of
three particles in an external field. The quantized motion of a particle in an
external field described in the symmetric gauge results in a set of infinitely
degenerate levels called Landau levels, with energy spacing between degenerate
manifolds of $\hbar\omega_{c}$. It is interesting to note that in
setting the interactions in Eq. \ref{Eq:HintHS} to zero, the Hamiltonian
becomes separable in hyperspherical coordinates, reproducing exactly the Landau
level picture, but with a slightly different interpretation of the level structure (discussed below).

The grand angular momentum operator is diagonalized using hyperspherical
harmonics \cite{smirnov1977method,avery1989hyperspherical} with eigenvalues
given by%
\begin{equation}
\Lambda^{2}Y_{\lambda m_{1}m_{2}}\left(  \Omega\right)  =\lambda\left(
\lambda+2\right)  Y_{\lambda m_{1}m_{2}}\left(  \Omega\right)
\end{equation}
where $\lambda$ is the grand angular momentum quantum number and $m_{1}$ and
$m_{2}$ are the 2D angular momenta associated with the Jacobi vectors
$\vec{\rho}_{1}$ and $\vec{\rho}_{2}$ respectively. Hyperspherical harmonics
also diagonalize the total angular momentum of the system as
\[
L_{z,int}Y_{\lambda m_{1}m_{2}}\left(  \Omega\right)  =MY_{\lambda m_{1}m_{2}%
}\left(  \Omega\right)
\]
where $M=m_{1}+m_{2}$ is the total 2D angular momentum of the system, The
allowed values of $\lambda$ are given by%
\begin{equation}
\lambda=2n+\left\vert m_{1}\right\vert +\left\vert m_{2}\right\vert
\label{Eq:Lambdarestrict}%
\end{equation}
where $n$ is a non-negative integer. Note that $\lambda$ has a minimum value
given by $\lambda=\left\vert m_{1}\right\vert +\left\vert m_{2}\right\vert $
when $n=0$.

Inserting the separability ansatz $\Psi\left(  \vec{\rho}_{1},\vec{\rho}%
_{2}\right)  =\Psi\left(  R,\Omega\right)  =R^{3/2}F\left(  R\right)
Y_{\lambda m_{1}m_{2}}\left(  \Omega\right)  $ into the Schr\"{o}dinger
equation resulting from Eq. \ref{Eq:HintHS} (with the interactions set to zero)
results in a 1D hyperradial Schr\"{o}dinger equation of a harmonic oscillator
with frequency $\omega_{c}/2$ that has been shifted down in energy
by $M\hbar\omega_{c}/2$, i.e.%
\begin{align}
0=  &  \left[  \dfrac{-\hbar^{2}}{2\mu}\dfrac{\partial^{2}}{\partial R^{2}%
}+\dfrac{\hbar^{2}}{2\mu}\dfrac{\left(  \lambda+1/2\right)  \left(
\lambda+3/2\right)  }{R^{2}}\right. \\
&  -\left.  \dfrac{\hbar\omega_{c}}{2}M+\dfrac{1}{8}\mu\omega_{c}^{2}%
R^{2}-E\right]  F\left(  R\right)  .\nonumber
\end{align}
Note that the $R^{3/2}$ factor in the separability ansatz is included to
remove first derivatives in the hyperradius. This Schr\"{o}dinger equation can
be solved simply \cite{rittenhouse2006hyperspherical} with eigenenergies and
eigenfunctions given by%
\begin{align}
E  &  =\hbar\omega_{c}\left[  \nu+\dfrac{\left(  \lambda-M\right)  }%
{2}+1\right]  ,v=0,1,2,...\label{Eq:NIEnergy}\\
F\left(  R\right)   &  =A_{\nu\lambda}\dfrac{e^{-R^{2}/(2\sqrt{2}l)_{c}}%
}{R^{3/2}}\left(  \dfrac{R}{\sqrt{2}l_{c}}\right)  ^{\lambda+3/2}L_{\nu
}^{\lambda+1}\left(  \dfrac{R^{2}}{2l_{c}^{2}}\right)  , \label{Eq:NIWF}%
\end{align}
where $l_{c}$ is the magnetic length, $L_{\nu}^{L}\left(  x\right)  $ is a
Lageurre polynomial and $A_{\nu\lambda}$ is a normalization constant$.$
Inserting the restriction on values of $\lambda$ from Eq.
\ref{Eq:Lambdarestrict} into Eq. \ref{Eq:NIEnergy} the Landau level picture
emerges:%
\begin{equation}
E=\hbar\omega_{c}\left(  \nu+n+\dfrac{\left\vert m_{1}\right\vert +\left\vert
m_{2}\right\vert -M}{2}+1\right)  . \label{Eq:LLenergies}%
\end{equation}
Restricting ourselves to positive values of angular momentum it is clear that
for fixed $\nu$ and $n$ any nonnegative value of total angular momentum $M$
produces the same energy, and thus an infinitely degenerate manifold of
states. The structure of the energy levels seen in Eq. \ref{Eq:LLenergies} is
the same as the energy levels seen in the standard Landau level
picture. Here, however, the interpretation of excitation between Landau levels
is somewhat different. There are two different ways to move from one level to
another, either through a hyperradial excitation by incrementing $\nu$, or
through a hyperangular excitation by incrementing $n$.

\subsection{Contact interactions and the adiabatic hypersperical method}

Next, we proceed to diagonalize the full interacting Hamiltonian of Eq.
\ref{Eq:HintHS} within the adiabatic hyperspherical method. The heart of the
approach is in treating the hyperradius $R$ as an adiabatic parameter, and
diagonaizing the Hamiltonian at fixed $R$ in the remaining hyperangular
degrees of freedom. In this method, the total wavefunction is expanded as
\begin{equation}
\Psi_{M}\left(  \vec{\rho}_{1},\vec{\rho}_{2}\right)  =\sum_{n}R^{3/2}%
F_{nM}\left(  R\right)  \Phi_{nM}\left(  R;\Omega\right)
,\label{Eq:Hyperspherexpand}%
\end{equation}
Here, the adiabatic channel functions, $\Phi_{nM}\left(  R;\Omega\right)  ,$
satisfy the fixed $R$ Schr\"{o}dinger equation%
\begin{equation}
\left[  \dfrac{-\hbar^{2}}{2\mu}\dfrac{\Lambda^{2}}{R^{2}}+\sum_{i<j}V\left(
r_{ij}\right)  \right]  \Phi_{nM}\left(  \Omega\right)  =u_{nM}\left(
R\right)  \Phi_{nM}\left(  \Omega\right)  ,\label{Eq:AdiabaticSE}%
\end{equation}
where $u_{nM}\left(  R\right)  $ is the adiabatic potential associated with
$\Phi_{nM}\left(  \Omega\right)  $. Note that this is exactly the adiabatic
Schr\"{o}dinger equation that is solved in finding the adiabatic potentials
for three bosons in the \emph{absence} of any external field. Thus
$u_{nM}\left(  R\right)  $ are simply the adiabatic potentials for three
interacting bosons confined to 2D, a system that has been studied extensively
\cite{nielsen1997three,nielsen2001three,petrov2000bose,jensen2004structure,kartavtsev2006universal,dincao2014adiabatic} and is of current interest in its own right.
Inserting Eq. \ref{Eq:HintHS} and projecting onto the $n$th channel function
results is a coupled system of one-dimensional Schr\"{o}dinger equations in
$R:$%
\begin{widetext}
\begin{align}
EF_{nM}\left(  R\right) = &  \left[  \dfrac{-\hbar^{2}}{2\mu}\dfrac{d^{2}%
}{dR^{2}}+U_{nM}\left(  R\right) - E \right]  F_{nM}\left(  R\right)
  -\dfrac{\hbar^{2}}{2\mu}\sum_{m\neq n}\left(  \mathbf{Q}_{nm}\left(
R\right)  +2\mathbf{P}_{nm}\left(  R\right)  \dfrac{d}{dR}\right)
F_{mM}\left(  R\right)  ,\label{Eq:CoupledSE}\\
U_{nM}\left(  R\right)  = &  u_{nM}\left(  R\right)  +\dfrac{\hbar^{2}}{2\mu
}\dfrac{3/4}{R^{2}}-\dfrac{\hbar^{2}}{2\mu}\mathbf{Q}_{nn}\left(  R\right)
  +\dfrac{1}{8}\mu\omega_{c}^{2}R^{2}-\dfrac{\hbar\omega_{c}}{2}M.\label{Eq:effectivePot}
\end{align}
\end{widetext}
Here the effective hyperradial potentials are given by $U_{n}\left(  R\right)
$ and the non-adiabatic corrections embodied in the $\mathbf{P}$ and
$\mathbf{Q}$ matrices are a result of hyperradial derivatives of the channel
functions, i.e.%
\begin{align}
\mathbf{P}_{nm} &  =\left\langle \left\langle \Phi_{nM}\left(  R;\Omega
\right)  \left\vert \dfrac{\partial}{\partial R}\Phi_{mM}\left(
R;\Omega\right)  \right.  \right\rangle \right\rangle ,\label{Eq:PMat}\\
\mathbf{Q}_{nm} &  =\left\langle \left\langle \Phi_{nM}\left(  R;\Omega
\right)  \left\vert \dfrac{\partial^{2}}{\partial R^{2}}\Phi_{mM}\left(
R;\Omega\right)  \right.  \right\rangle \right\rangle ,\label{Eq:Qmat}%
\end{align}
where the double bracket $\left\langle \left\langle {\cdot}\right\rangle
\right\rangle $ indicates that the matrix elements are taken over the
hyperangular degrees of freedom only.

\begin{figure}[tbh]
\begin{center}
\includegraphics[width=3in]{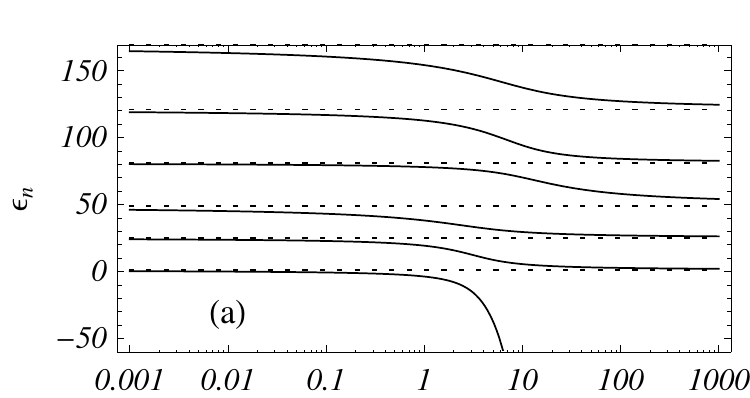}
\includegraphics[width=3in]{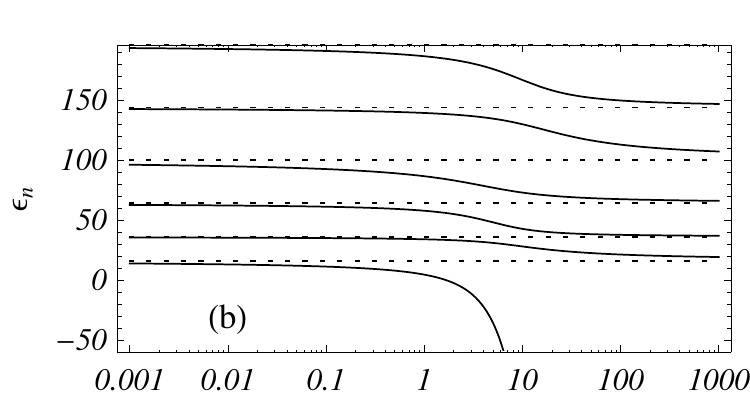}
\includegraphics[width=3in]{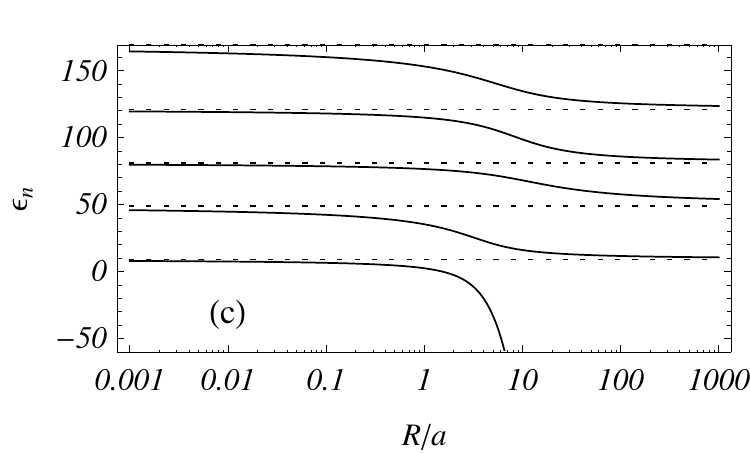}
\end{center}
\caption{The first several adiabatic hyperangular eigenvalues for angular
momentum (a) $M=0$, (b) $M=1$, and (c) $M=3$ are shown as a function of $R/a$
on a log scale. Dotted lines indicate the hyperangular eigenvalues of
hyperspherical harmonics of non-interacting systems expected in the large and
small $a$ limits.}%
\label{Fig:Hyperangvals}%
\end{figure}

Up to this point the treatment described above has been quite general, and is applicable to any two-body, cylindrically symmetric interaction. In fact, by extending the Jacobi coordinates to larger
numbers of particles, this treatment can be extended to any $N$-body system.
The adiabatic Schr\"{o}dinger equation (\ref{Eq:AdiabaticSE}) has been
solved for three-body systems with several different interaction potentials
 and can be approached by a number of different techniques \cite{nielsen1999structure,hammer2004universal,kartavtsev2006universal,dincao2014adiabatic}.
In this work we focus on the zero-range pseudo-potential \cite{kanjilal2006coupled,shih2009pseudopotential},%
\begin{equation}
V\left(  r\right)  =\dfrac{\hbar^{2}}{m}\dfrac{\delta\left(  r\right)
}{r\left[  1-\ln\left(  \dfrac{r}{a}\right)  \right]  }\dfrac{\partial
}{\partial r}r. \label{Eq:Pseudopot}%
\end{equation}
where $a>0$ is the 2D, s-wave ($m=0$) scattering length. The effect of this
pseudo-potential is to enforce the two-body boundary condition
\begin{align}
\lim_{r_{ij}\rightarrow0}\dfrac{\partial}{\partial r_{ij}}\Phi_{nM}\left(
R;\Omega\right)   &  =C\left[  1+\dfrac{2\tan\delta}{\pi}\left(  \ln
\dfrac{kr_{ij}}{2}+\gamma\right)  \right]  ,\label{Eq:BC}\\
\tan\delta &  =\dfrac{\pi}{2\left(  \ln\dfrac{ka}{2}+\gamma\right)
}.\nonumber
\end{align}
where $C$ is a constant, $\delta$ is the 2D s-wave scattering phase shift, $k$
is the two-body wavenumber, and $\gamma=0.5772$ is the Euler constant. It is
important to note that this pseudo-potential only affects wavefunctions with
an s-wave component of the angular momentum between pairs of particles; all
higher partial waves are treated as non-interacting. This pseudo-potential can
be used when the true inter-particle interaction falls off sufficiently fast
at large $r$ to be considered short-range--the scattering length is much
larger than both the range, $r_0$, and the effective range, $r_{eff}$, of the interaction, i.e. $a\gg r_{0}$ and $a\gg r_{eff}$. At the
two-body level this pseudo-potential interaction produces a large halo dimer
state with binding energy $E_{b}=-4e^{-2\gamma}/ma^{2}$ \cite{shih2009pseudopotential}. It is also worth
noting that in the limit of large or small scattering length, $ka\rightarrow
\infty$ \emph{or }$ka\rightarrow0$, where $k$ is the relative two-body
momentum, the pseudo-potential approaches the non-interacting limit.

In the present study, we employ the hyperangular Green's function approach of Ref. \cite{rittenhouse2010greens}. The full
derivation using this method is somewhat tedious, but straightforward, and we
will not detail it here. The heart of the method is in turning the adiabatic
Schr\"{o}dinger equation into a Lippmann-Schwinger (LS) equation by employing
the free-space hyperangular Green's function. Within this LS equation, it is
easy to apply the boundary condition of Eq. \ref{Eq:BC} when two particles are
in contact with each other, and propagate the particles freely between
such contact points. The result of this derivation gives the adiabatic
potentials as
\begin{equation}
u_{nM}\left(  R\right)  =\dfrac{\hbar^{2}}{2\mu}\dfrac{\varepsilon_{nM}\left(
R\right)  -1}{R^{2}}.
\end{equation}
where we refer to $\varepsilon_{nM}\left(  R\right)  $ as the hyperangular
eigenvalues which are roots of the transcendental equation \begin{widetext}
\begin{align}
-\ln\dfrac{R}{a}=  & \ln\sqrt{\dfrac{2\mu}{m}}-\gamma-\dfrac{1}{2}\psi\left(
\dfrac{M-\sqrt{\varepsilon_{nM}}+1}{2}\right)  -\dfrac{1}{2}\psi\left(
\dfrac{M+\sqrt{\varepsilon_{nM}}+1}{2}\right)  \label{Eq:Trancend}\\
& +\left(  \dfrac{-1}{2}\right)  ^{M}\dfrac{\Gamma\left(  \dfrac
{M+\sqrt{\varepsilon_{nM}}+1}{2}\right)  \Gamma\left(  \dfrac{M-\sqrt
{\varepsilon_{nM}}+1}{2}\right)  }{\Gamma\left(  M+1\right)  }{}_{2}F_{1}\left(
\dfrac{M-\sqrt{\varepsilon_{nM}}+1}{2},\dfrac{M+\sqrt{\varepsilon_{nM}}+1}%
{2};M+1;\dfrac{1}{4}\right)  ,\nonumber
\end{align}
where $\Gamma\left(  x\right)  $ is the gamma function, $\psi\left(
x\right)  =\Gamma^{\prime}\left(  x\right)  /\Gamma\left(  x\right)  $ is the
digamma function and ${}_{2}F_{1}\left(a,b;c;x\right)$ is a hypergeometric function. In the case where
$M=0$, this reduces to the results found in Refs. \cite{nielsen1997three,kartavtsev2006universal}.
\end{widetext}

Examples of the hyperradial eigenvalues $\varepsilon_{nM}$ for $M=0,1$ and
$3$ are shown in Fig. \ref{Fig:Hyperangvals} as a function of $R/a$. In each
case the lowest hyperangular eigenvalue goes to $-\infty$ quadratically in $R$
in the large $R$ limit. This corresponds to a particle-dimer hyperangular
channel function consisting of a free particle far away from a bound dimer. With
the exception of the lowest potential in the large $R$ limit, all hyperangular
eigenvalues logarithmically approach an integer, corresponding to a
non-interacting value, in both the large and small $R$ limits. The
hyperangular eigenvalues transition from one non-interacting limit to another
in the region where $R\sim a$.

We can understand this behavior by considering the pseudo-potential in Eq.
\ref{Eq:Pseudopot}. In the limit of large hyperradius, $R\gg a$, the average
inter-particle separation is much greater than the scattering length, $r\gg
a.$ The logarithmic behavior of the scattering length in the pseudo potential
indicates that this is a weakly repulsive limit. In the limit of very small
hyperradius, $R\ll a$, the average inter-particle separation is much smaller
than the scattering length, $r\ll a$, again, because of the logarithmic nature
of the pseudo-potential, this becomes the weakly attractive limit.

The matrix elements of the non-adiabatic correction matrix, $\mathbf{P}$ (eq (\ref{Eq:PMat}))are given
in Ref. \cite{kartavtsev2006universal} in terms of the hyperangular eigenvalues by%
\begin{equation}
\mathbf{P}_{mn}=\dfrac{\sqrt{\varepsilon_{mM}^{\prime}\left(  R\right)
\varepsilon_{nM}^{\prime}\left(  R\right)  }}{\varepsilon_{mM}\left(
R\right)  -\varepsilon_{nM}\left(  R\right)  } \label{Eq:Pmat}%
\end{equation}
where the primes indicate a derivative with respect to $R$. The diagonal
correction, $\mathbf{Q}_{nn}\left(  R\right)  $, to the potentials in Eq.
\ref{Eq:effectivePot} is given by
\begin{equation}
\mathbf{Q}_{nn}=-\dfrac{1}{12R^{2}}-\dfrac{1}{4}\left(  \dfrac{\varepsilon
_{nM}^{\prime\prime}\left(  R\right)  }{\varepsilon_{nM}^{\prime}\left(
R\right)  }\right)  ^{2}+\dfrac{\varepsilon_{nM}^{\prime\prime\prime}\left(
R\right)  }{6\varepsilon_{nM}^{\prime}\left(  R\right)  }. \label{Eq:Qdiag}%
\end{equation}
In the infinite channel limit, the off-diagonal elements of $\mathbf{Q}$ can
be found using the identity%
\begin{equation}
\mathbf{Q}_{mn}=\mathbf{P}_{mn}^{\prime}\left(  R\right)  +\left[
\mathbf{P}^{2}\right]  _{mn} \label{Eq:Qoffdiag}%
\end{equation}
where $\left[  \mathbf{P}^{2}\right]  _{mn}$ is the $mn^{th}$ element of the
square of the P-matrix and $\mathbf{P}^{\prime}$ indicates a derivative with respect to the hyperradius. While Eq. \ref{Eq:Qoffdiag} is only exact in the
infinite channel limit, for the purposes of this work, we will use it in a
finite channel number expansion to approximate direct, off-diagonal,
non-adiabatic contributions.

\section{Adiabatic potentials}

\begin{figure}[tbh]
\begin{center}
\includegraphics[width=3in]{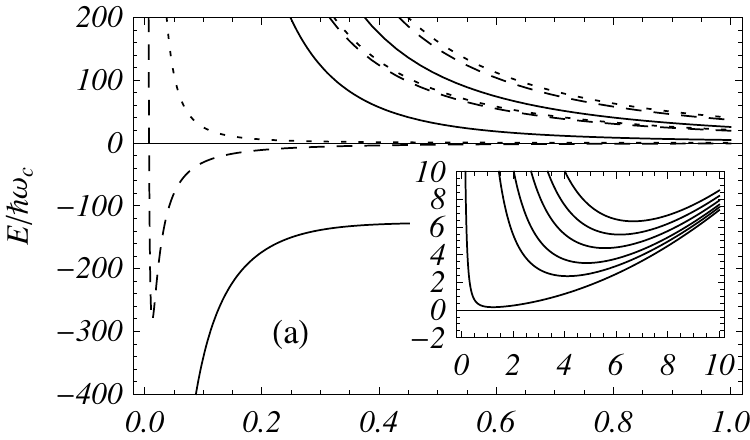}
\includegraphics[width=3in]{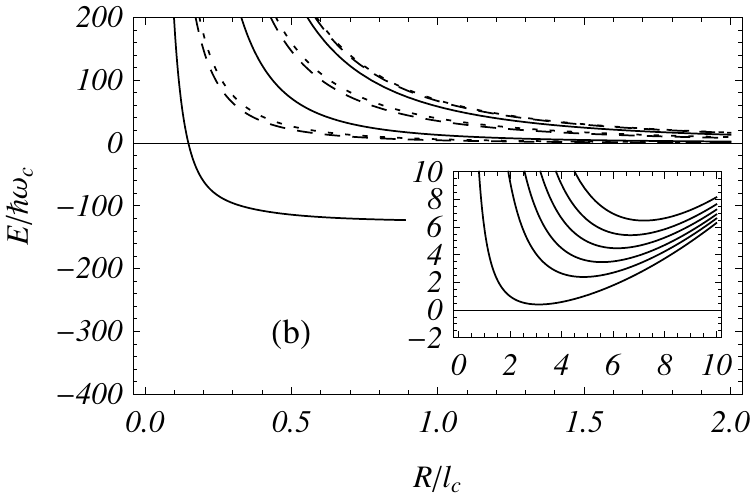}
\end{center}
\caption{The lowest three adiabatic potentials are shown for $M=0$ (a) and
$M=2$ (b) as a function of $R$ in units of the magnetic length $l_{c}$ for
several different value of the scattering length, $a=0.1l_{c}$ (solid curves),
$a=7l_{c}$ (dashed curves) and $a=10,000l_{c}$ (dotted curves). (Insets) The
lowest three adiabatic potentials for $a=10,000l_{c}$ are shown over a
different scale to illustrate the effective oscillator potential imposed by
the external magnetic field.}%
\label{Fig:Potentials}%
\end{figure}

One of the strengths of the adiabatic hyperspherical method is that, with the
effective adiabatic potentials in hand, we can bring all of the understanding
and intuition of normal 1D Schr\"{o}dinger quantum mechanics to bear on the
problem. The adiabatic potentials of Eq. \ref{Eq:effectivePot} can give
significant insight into the structure and behavior of the three-body system.
Figure \ref{Fig:Potentials}a shows the lowest three potentials for total angular
momentum $M=0$ for scattering lengths $a=0.1l_{c},$ $a=7 l_{c}$ and $a=10,000 l_{c}%
$. Figure \ref{Fig:Potentials}b show the same for a system with total angular
momentum $M=2$. The insets in Figs. \ref{Fig:Potentials}a and b show the
potentials for $a=10000l_c$ over a larger range of $R$ to illustrate the effective
oscillator potential that is introduced as a consequence of the magnetic
field. In the adiabatic potentials, there is a competition between two length scales;
the scattering length, $a$, which controls the interactions, and the magnetic
length, $l_{c}$, which is controlled by the magnetic field. In the small $a$
limit, the minimum of lowest adiabatic potential is shifted down far below the zero
energy threshold. In fact in the absence of any field this potential
asymptotically goes to an energy of $U\rightarrow-4\exp\left(  -2\gamma
\right)  \hbar^{2}/ma^{2}$ which is exactly the energy of the dimer state. As
a result, we can interpret the lowest potential in the small scattering length
limit as describing an effective two-body interaction potential of a particle
and an $m=0$ dimer in a weak external field, a behavior that persists for all
values of $M$.

In the small scattering length limit the second adiabatic potential is similar
to a harmonic oscillator potential, and in the absence of any magnetic field
asymptotes to the zero-energy threshold. This allows us to interpret the
lowest adiabatic interaction channel as that corresponding to the behavior of
a three-body system with no dimer type bound states. Another feature that can
be seen in the lowest $M=0$ adiabatic potential is a short range attractive
well that is not present for any other values of $M$. For small $a$ this well
is deep enough to bind two three-body states with binding energies of
$E=16.25E_{b}$ and $1.26E_{b}$ respectively, where $E_{b}$ is the dimer
binding energy. These values are found using a single channel calculation and
are in good agreement with three-body bound state energies in the absence of
the magnetic field found in Refs. \cite{kartavtsev2006universal,helfrich2011resonant}.

In the large scattering length limit, $a\gg l_{c}$, the hyperangular
eigenvalues are dominated by the small hyperradius behavior, meaning that the
potentials are close to the noninteracting limit. In this limit the particles
are all kept close together by the (strong) external field and the average
inter-particle separation is much smaller than the scattering length. This
means that any particle-dimer type behavior is pushed to the the large $R$
regime energetically far removed from the minimum. Thus all of
the potentials can be considered to correspond with true three-body behavior.

In general when the hyperradius is much larger than the scattering length
$R\gg a$ , the three-body potentials (those not associated with
particle-dimer type behavior) approach the non-interacting limit. When the
scattering length is much smaller than the magnetic length, the minimum in the
harmonic potential resides at $R\sim l_{c}\gg a$ and deviations from the non-interacting behavior are
energetically inaccessible.
Therefore, we can expect the system
to approximately behave as three non-interacting particles in an external field. For the
lowest potential, associated with particle-dimer behavior, with the
exception of the $M=0$ states, we can expect that the system will behave as a
non-interacting two-body system in an external field whose ground state energy
is shifted by the dimer binding energy. For $M=0$ the deep well in the $R<a$
region in the potentials will modify this behavior.

 In the $R\ll a$ limit, all of the potentials have non-interacting limiting behavior. When the scattering
length is much smaller that the magnetic length, any changes in the potentials
that result from the interaction are in the small hyperradius region, pushed
far up the inner potential barrier that can be seen in the inset of Fig.
\ref{Fig:Potentials}(a,b). We can therefore expect that the behavior of the
system will again approach that of the non-interacting system in the $a\gg
l_{c}$ limit. With the potentials in hand, we can now examine the
eigenspectrum of the system.

\section{Three-body energies}

In this section we describe the behavior of the eigenspectrum that results
from the coupled system of 1D Schr\"{o}dinger equations from Eq.
\ref{Eq:CoupledSE}. Generally, it is necessary
to include many adiabatic channels to converge the bound state energies of
three-bodies in two dimensions to high accuracy; however, to achieve accuracy
to within several digits, only relatively few channels are needed. With that
in mind, we solve Eq. \ref{Eq:CoupledSE} using the lowest six adiabatic
channels for each total angular momentum $M$. We have found that this is
sufficient to converge the energetically low lying states to within
$\sim0.1\%$ accuracy, which is sufficient for the purposes of this study.
\begin{figure}[tbh]
\begin{center}
\includegraphics[width=3in]{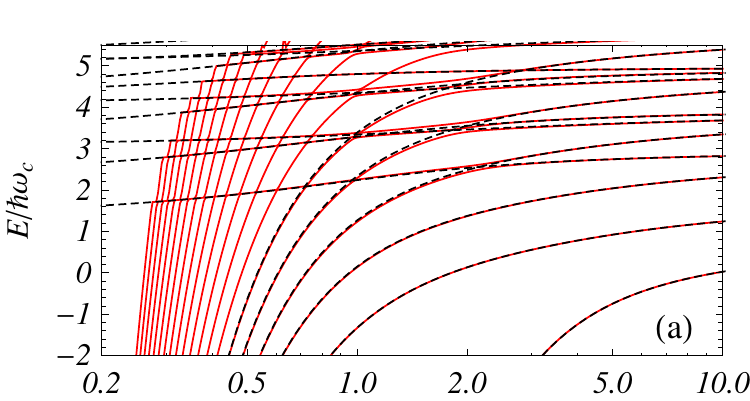}
\includegraphics[width=3in]{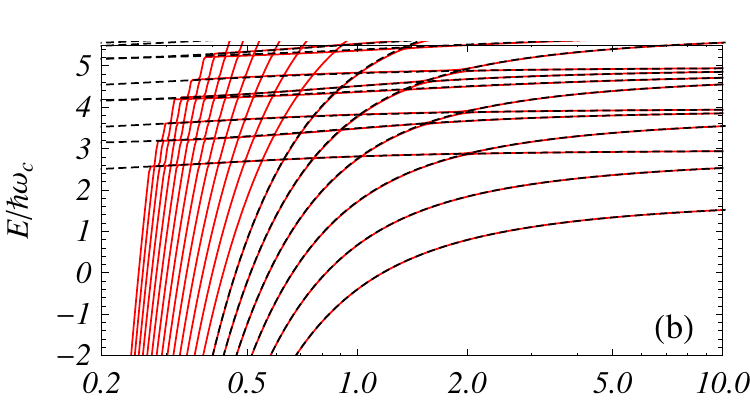}
\includegraphics[width=3in]{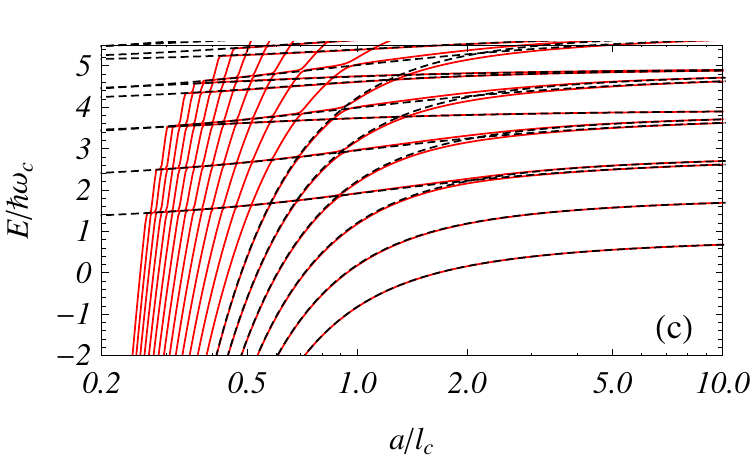}
\end{center}
\caption{(color online) The full energy spectrum in units of $\hbar\omega_{c}$
is shown as a function of $a/l_{c}$. The energies were calculated for total
angular momentum $M=0$ (a), $M=1$ (b), and $M=2$ (c) including (solid red) and
ignoring (dashed black) couplings between adiabatic channels . }%
\label{Fig:FullEnergies}%
\end{figure}

The full eigenspectrum of the three-boson system for $M=0$, $1$, and $2$
including the off-diagonal couplings between adiabatic channels is shown as a
function of $a/l_{c}$ in red in Fig.~\ref{Fig:FullEnergies}(a-c). Also shown
in Fig.~\ref{Fig:FullEnergies}(a-c) are the energies of the lowest six hyperradial vibrational states for
each effective potential, $U_{nM}\left(  R\right)  $, ignoring the off
diagonal coupling matrices $\mathbf{P}$ and $\mathbf{Q}$ but including the
diagonal correction $\mathbf{Q}_{nn}(R)$. In regions nearby crossings between
uncoupled energies, the off diagonal couplings $\mathbf{P}_{nm}(R)$ and
$\mathbf{Q}_{nm}(R)$ introduce a series of narrow avoided crossings in the
energy spectrum as a function of the scattering length and these off diagonal
direct and derivative couplings become important in these areas. However, away
from these avoided crossings, the uncoupled adiabatic potentials
$U_{nM}\left(  R\right)  $ provide a good approximation of the energy spectrum
of the three-particles system indicating that the system is nearly separable
within the adiabatic hyperspherical framework which provides an accurate
description of the this system. As a result, unless otherwise stated, we will
focus on the uncoupled adiabatic channel energies for the remainder of this paper.

\begin{figure}[tbh]
\begin{center}
\includegraphics[width=3in]{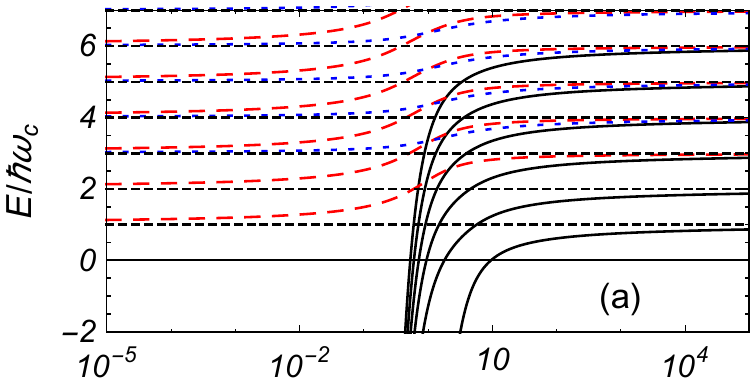}
\includegraphics[width=3in]{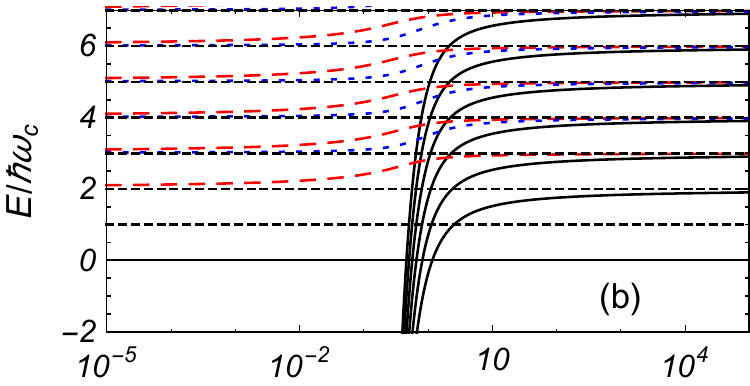}
\includegraphics[width=3in]{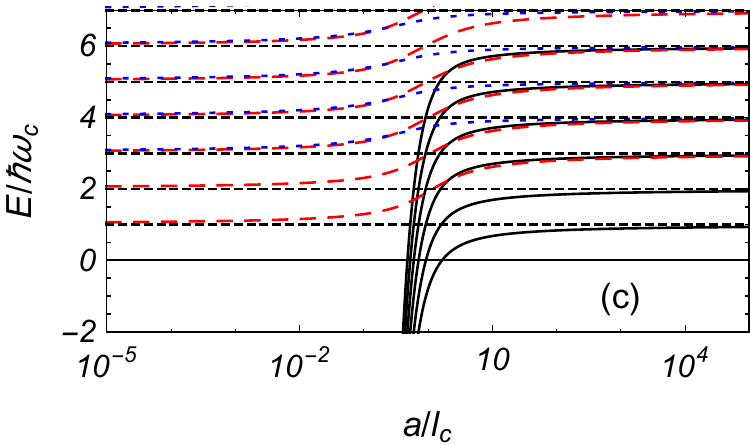}
\end{center}
\caption{(color online) The energies of the first six vibrational levels attached to
the first (solid black curves), second (dashed red curves), and third (dotted
blue curves) adiabatic channels are shown as a function of scattering length
for internal angular momentum (a) $M=0$, (b) $M=1$, and (c) $M=2$. The
coupling between adiabatic channels has been ignored here. The Landau Level
energies are shown as black dashed lines for reference.}%
\label{Fig:UncoupledEns}%
\end{figure}


Figures \ref{Fig:UncoupledEns}(a-c) show the vibrational energies for the
lowest three adiabatic channels as a function of $a/l_{c}$ for $M=0,1,$ and
$2$ respectively. In each case, the lowest adiabatic channel has vibrational
states that decrease in energy as $1/a^{2}$ for $a\ll l_{c}$. These are states
associated with an atom-dimer interaction channel. For $M=0$ the lowest two
vibrational states in the lowest channel become the three-body bound
state mentioned previously. While atom-dimer states are of interest in their
own right, we are interested here in the behavior of three-body states in the
presence of an external field and the atom-dimer states in the small
scattering length limit will not be the focus of this work. For the second and
third adiabatic channels, we can see that at very small scattering lengths
the three-body energies approach the noninteracting Landau level values as
expected. In sweeping from small to large scattering length the energies
transition up smoothly to a higher landau level in the large scattering length
limit. In the small $a$ limit the energies are shifted up slightly from the
non-interacting energy corresponding to an effectively repulsive interaction.
The energies of the system in the large scattering length limit are
shifted slightly down from the noninteracting Landau levels corresponding to an effectively attractive interaction.

In the large and small scattering length limits Fig. \ref{Fig:UncoupledEns}b
shows that the lowest $M=1$ adiabatic channel that corresponds to a three-body
state converges to the second Landau level rather than the first. As discussed
later, this is because a total internal angular momentum of $M=1$ in the
lowest Landau level is forbidden for bosonic symmetry and the lowest
non-interacting three-boson state with internal angular momentum $M=1$
corresponds to the second Landau level.

\begin{figure}[tbh]
\begin{center}
\includegraphics[width=3in]{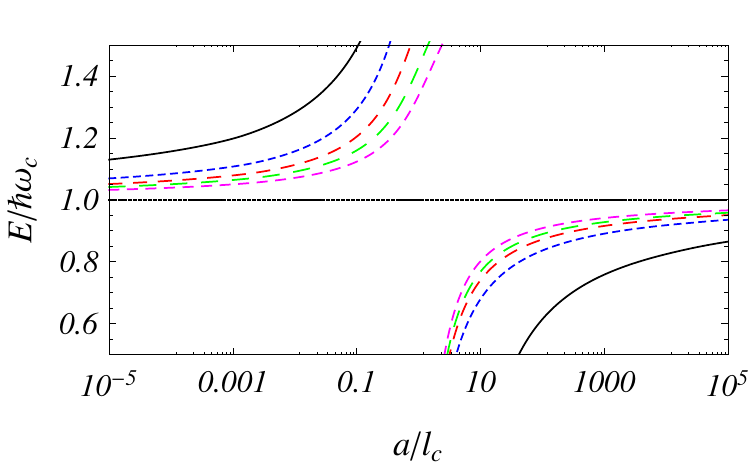}
\end{center}
\caption{(color online) The lowest vibrational state energy for each adiabatic
channel is shown for $M=0,2,3,4$ and $5$ as a function of $a/l_{c}$ on a log scale. The energies for $M=0$
are shown as solid curves while the energies for $M=2-5$ are shown as
dashed curves with increasing dash size for increasing values of $M$. The
non-interacting Landau level is shown for reference.}%
\label{Fig:LowestVibEns}%
\end{figure}

Of particular interest here is the behavior of the lowest energy three-body
state. Figure \ref{Fig:LowestVibEns} shows the energy of the lowest three-body
state for $M=0$, $2$, $3$, $4$, and $5$ (shown in black, blue, magenta, red,
and green respectively--color online) as a function of $a/l_{c}$ in the region near the
lowest Landau level. Because there are no $M=1$ lowest Landau level states, $M=1$
is excluded here. There are several interesting things that can be observed in
this figure. First, we can see that at small scattering length, $a<l_{c}$, the
three-body energies are pushed above the lowest Landau level indicating that
this interaction regime corresponds to repulsive interactions. For large
scattering length, $a>l_{c}$, the three-body energies are below the lowest
Landau level indicating that this is the attractive interaction regime. We
also note that for $M\geq2$ the levels show a parity oscillation, with $M$
even higher in energy for small scattering length and $M$ odd
lower in energy, and vice-versa in the large scattering length limit. 

\begin{figure}[tbh]
\begin{center}
\includegraphics[width=3in]{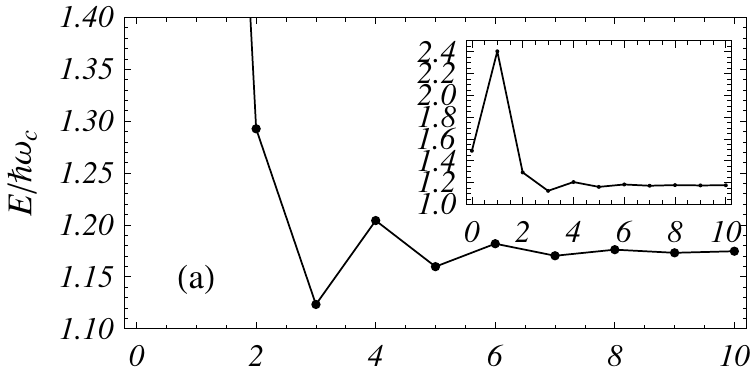}
\includegraphics[width=3in]{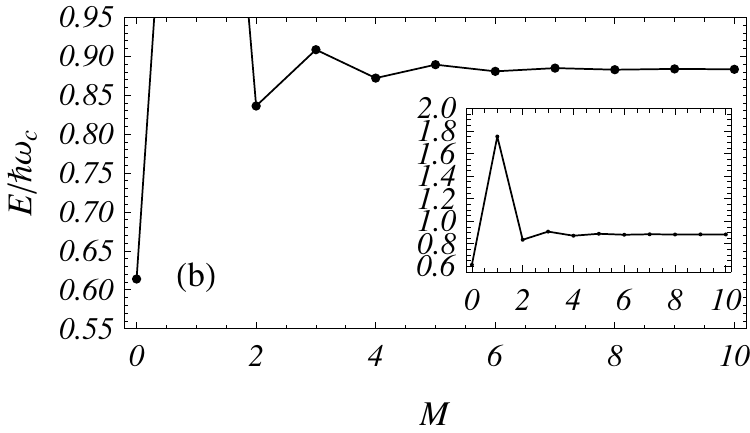}
\end{center}
\caption{The energy of the lowest three-body state (the state without any
dimer-like characteristics) is shown as a function of internal angular
momentum $M$ for (a) the weakly repulsive ($a=0.1l_{c}$) and (b) weakly attractive ($a=100l_{c}$)
regimes. Note that for $M>1$ the energies oscillate
with parity with odd $M$ being lower in energy for small scattering length and even
$M$ being lower for large scattering length. Universally for small scattering
length, the lowest $M=3$ state is the lowest interacting three-body state for
small $a/l_{c}$. (Insets) Shows the same states on a larger scale to include
the lowest $M=1$ state.}%
\label{Fig:LowestE}%
\end{figure}

The opposite parity oscillation is shown in Fig. \ref{Fig:LowestE} in which we
have plotted the lowest energy three-body state versus internal angular
momentum $M$ for small scattering length ($a=0.1l_{c}$ in Fig.
\ref{Fig:LowestE}a) and large scattering length ($a=100 l_{c}$ in Fig.
\ref{Fig:LowestE}b). This parity oscillation can be understood simply by
understanding that even parity states (even $M$) tend to have the three bosons
in closer proximity and thus more strongly feel the s-wave contact
interactions than the odd parity (odd $M$) states. It is interesting to note
here that because $M=1$ is forbidden for bosonic symmetry in the lowest Landau
level, $M=3$ is \emph{universally} the lowest energy three-body state for
states interacting via repulsive, $a<l_{c}$, s-wave, contact interactions.

At larger angular momentum, we would expect an angular momentum (centrifugal) barrier to
form that prevents the bosons from being near each other. Thus we might expect
that the energy should tend towards the non-interacting value of the lowest
Landau level. This behavior is not born out in Figure \ref{Fig:LowestE}, where
we observe that the energies of the three-boson states tend towards a
constant that is above the lowest Landau level for repulsive interactions and
below it when the interactions are attractive. The explanation for this
apparent inconsistency lies in the regularized s-wave contact interaction of
Eq. \ref{Eq:Pseudopot}. This interaction projects onto only those states which
have some component of their wavefunctions with zero inter-particle angular
momentum. Examining the energy level structure of the lowest Landau level from Eq. \ref{Eq:LLenergies} with $m_1 = 0$, we can see that for each value of total internal angular momentum,
there is at most one such lowest Landau level state. When the lowest Landau
level for a given value of $M$ is degenerate, there will be additional states
in which the inter-particle angular momentum has no $m=0$ component and will
be non-interacting according to our pseudo-potential.

\section{The role of degeneracy}

Recently, Daily, \emph{et al.} \cite{daily2015hyperspherical} have highlighted the
crucial role that the degeneracy of the lowest Landau level plays in the
energetic structure of 2D few-fermion systems interacting via the Coulomb
potential in the presence of an external magnetic field. This degeneracy plays
an equally important role in the present study. As mentioned above, at
most only a single lowest Landau level state for each internal angular momentum
value, $M$, has a zero inter-particle angular momentum component in the
three-boson system. However, as the value of $M$ increases the degeneracy
generally increases as well. A complete description of the degeneracy of the
lowest Landau level for $N$ identical fermions (which can be directly applied
to $N$ identical bosons) is given in Ref. \cite{daily2015hyperspherical}, we will briefly reiterate the
argument here for completeness.

The non-interacting $N$-body Hamiltonian from Eq. \ref{Eq:TotHam}, with
$V\left(  r\right)  =0$, is separable in individual particle coordinates
where the bosonic lowest Landau level wavefunctions are given by%
\begin{equation}
\Psi_{LL}=\mathcal{N}\mathbf{\hat{S}}\prod_{j=1}^{N}z_{j}^{m_{j}%
}e^{-\left\vert z_{j}\right\vert ^{2}/l_{c}^{2}}, \label{Eq:IndepPartLLL}%
\end{equation}
where $z_{j}=x_{j}+iy_{j}$ is the $j$th boson's position in the laboratory frame written in complex
coordinates and $m_{j }\ge 0$ is the angular momentum of particle $j$. Here,
$\mathbf{\hat{S}}$ is a symmetrization operator that imposes bosonic symmetry and
$\mathcal{N}$ is a normalization factor. In the symmetrized basis the
individual angular momenta are no longer good quantum numbers; however, the
total angular momentum, $M_{Tot}=\sum_{j=1}^{N}m_{j}$, is conserved. Using
this the degeneracy of the lowest Landau level is given by the number of ways
we can combine $N$ individual particle angular momenta to get $M_{Tot}$
subject to the condition that $m_{1}\leq m_{2}\leq m_{3}...$ so that we do not
double count any symmetric configurations. This means that the degeneracy,
$D_{Tot}\left(  M_{Tot}\right)  $ for total angular momentum, $M_{Tot }%
$, for $N$ particles is given by the number of integer partitions, $P_N$, of no more
than $N$ integers of $M_{Tot},$ i.e. $D_{Tot}=P_{N}\left(  M_{Tot}\right)  $
where we define $P_{N}\left(  0\right)  \equiv 1$.

The non-interacting Hamiltonian is also separable into internal and center of
mass degrees of freedom as seen in Eq. \ref{Eq:Hcmsep}. In this basis the
total angular momentum can be written in terms of the internal angular
momentum, $M$, and the center of mass angular momentum, $M_{CM}$ as
$M_{Tot}=M+M_{CM}$. The degeneracy of the lowest Landau level for internal
angular momentum $M$, $D\left(  M\right)  $, is given by the degeneracy when
the total angular momentum is entirely internal, i.e. $M_{Tot}=M$ and
$M_{CM}=0$. However, the degeneracy, $D_{Tot}\left(  M_{Tot}\right)  ,$ from
above includes all allowed values of the center of mass angular momentum,
$0\leq M_{CM}\leq M_{Tot}$. To find only the degeneracy of the states with
$M_{CM}=0$, we must subtract off the total number of configurations with
$M_{CM}=1,2,3,...,M$. Because each center of mass angular momentum in
non-degenerate, and the fact that for each value of $M$ there is only one value of $M_{CM}$ that
gives $M_{Tot}=M+M_{CM}$, the number of states with $M_{CM}=1,2,3,...,M$ with
$M_{Tot}=M$ is given by the number of states with $M_{Tot}=M-1$, i.e.%
\begin{align}
D\left(  M\right)   &  =D_{Tot}\left(  M\right)  -D_{Tot}\left(  M-1\right)
,\label{Eq:Degeneracy}\\
&  =P_{N}\left(  M\right)  -P_{N}\left(  M-1\right)  ,\nonumber
\end{align}
with $P_N\left(  -1\right)  \equiv 0$.

\begin{figure}[tbh]
\begin{center}
\includegraphics[width=3in]{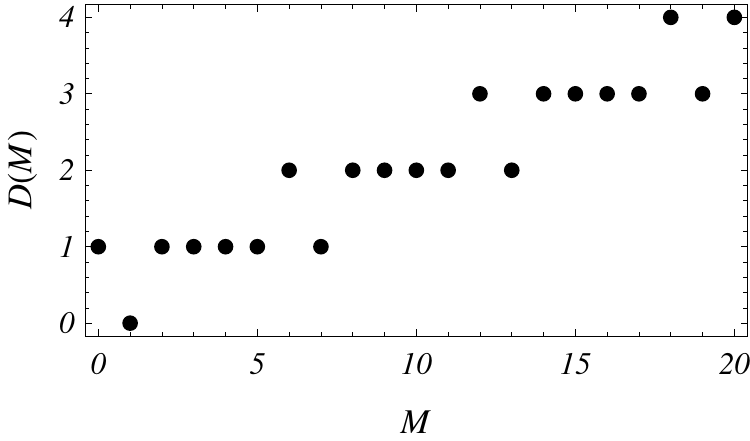}
\end{center}
\caption{The degeneracy of the lowest Landau level for three identical bosons
is shown as a function of internal angular momentum $M$. Note that for $M=6n$
where $n=1,2,3,...$ there is an anomalously high level of degeneracy.}%
\label{Fig:Degeneracy}%
\end{figure}

Figure \ref{Fig:Degeneracy} shows the degeneracy of the lowest Landau level
for three identical bosons as a function of internal angular momentum $M$.
Several interesting things emerge in this figure. First, we can see that, as
stated above, there are no allowed bosonic states with internal angular
momentum $M=1$. This is because there is only one lowest Landau level state
with total angular momentum $M_{Tot}=1$ when $m_{1}=0$, $m_{2}=0$, and
$m_{3}=1$ where $m_{i}$ is defined as in Eq. \ref{Eq:IndepPartLLL}. Since we
know there is a bosonic state with internal angular momentum $M=0$, the only
way to get total angular momentum $M_{Tot}=1$ is then with $M=0$ and
$M_{CM}=1$.

We can also observe in Fig. \ref{Fig:Degeneracy} that for $M=6n$,
$n=1,2,3,...$ the lowest Landau level presents an unusually high level of
degeneracy where the degeneracy is higher than both that of $M+1$ and $M-1$.
We note that these anomalously high degeneracy values of $M$ correspond
exactly to the angular momentum of the three-boson Laughlin states \cite{laughlin1} whose
wavefunctions are given by%
\begin{equation}
\Psi_{L}\left(  z_{1},z_{2},z_{3}\right)  =\mathcal{N}e^{\left(  -\sum
_{j}\left\vert z_{j}\right\vert ^{2}/l_{c}^{2}\right)  }\left[  \prod
_{i<j}\left(  z_{i}-z_{j}\right)  ^{2n}\right]  , \label{Eq:Laughlin}%
\end{equation}
where $\mathcal{N}$ is again a normalization constant and $n=1,2,3,...$. These
states are lowest Landau level states in which each particle pair has an
inter-particle angular momentum of $2n$ giving a total angular momentum of
$M_{Tot}=N\left(  N-1\right)  n=6n$ for $N=3$. Because the prefactor in Eq.
\ref{Eq:Laughlin} only depends of the position of the particles relative to
each other, it contains no center of mass angular momentum, $M_{CM}=0$. We
also note that the same degeneracy pattern as seen in Fig.
\ref{Fig:Degeneracy} for three bosons occurs for three identical fermions, but
shifted to the right by $M=3$ (the minimum allowed angular momentum for three
fermions in the lowest Landau level).

As discussed above, when the lowest Landau level becomes degenerate, for $M=6$
and $M\ge 8$ here, there exists at least one state in which each particle pair
has no zero angular momentum component, and which consequently do not
experience the s-wave pseudo potential. This means that, for weakly repulsive
interactions, $a\ll l_{c}$, the lowest energy three-body states are
non-interacting lowest Landau level states with $E=\hbar\omega_{c}$ for these
values of $M$.

If a repulsive d-wave interaction were to be included in this system, any
state with an $m=2$ component in its inter-particle angular momentum would
experience the interaction. We can assume that the d-wave interaction would
generally have a smaller effect on these states than the s-wave interaction
has on states with an $m=0$ inter-particle angular momentum component.
However, we surmise that the pattern of the energy shift from the lowest
Landau level would be similar for these d-wave interacting states to the
s-wave interacting states, mainly the pattern of even/odd parity oscillation.
Further, when the degeneracy jumps up to 3 degenerate states or more (for
$M=12$ and $M\geq14$) there will exist states with only $m\geq4$
inter-particle angular momentum components. These higher angular momentum
states will not experience either the s- or d-wave interactions.

\begin{figure}[tbh]
\begin{center}
\includegraphics[width=3in]{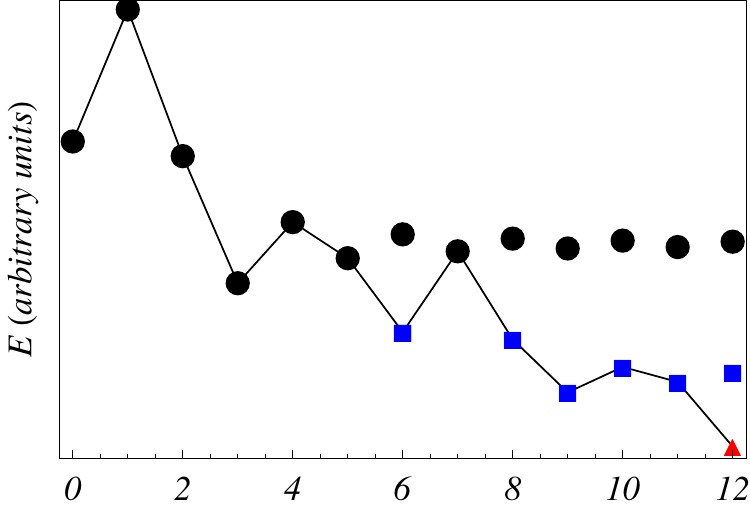}
\end{center}
\caption{(color online) A schematic illustration of the possible energy structure of the
three-boson system is shown as a function of internal angular momentum $M$.
This schematic illustrates the possible interplay between even/odd parity
oscillations in the energy and the degeneracy pattern which could result in
ground state energies that reflect a ``magic number'' type behavior.}%
\label{Fig:Enschem}%
\end{figure}

Figure \ref{Fig:Enschem} shows a schematic representation of the energies that
might result from including a repulsive d-wave interaction.  We emphasize here that the energies shown here are purely schematic in nature and are shown only to illustrate the surmised structure of the lowest energy states for weakly repulsive interactions. Here, black
circles represent the energies of states which are experiencing the s-wave
interaction, while blue squares represent states which only experience a
d-wave interaction. A singe red triangle at $M=12$ represents a state with no
s- or d-wave inter-particle angular momentum (this is in fact the Laughlin
state of Eq. \ref{Eq:Laughlin} with $n=2$). The even/odd oscillation combined
with the pattern of degeneracies for three identical bosons creates an
interesting overall pattern of lowest energy states (marked in Fig.~\ref{Fig:Enschem} by the
solid line) in which the ground state energy for each value of $M$ tends to
decrease overall with increasing $M.$ However, every third state, $M=0,3,6,9,$
and $12$ here, is lower in energy than either of its neighbors. This pattern
of anomalously low energy states is similar to the \textquotedblleft magic
number\textquotedblright states first predicted in Ref. \cite{girvin1983interacting}
for fermions interacting via the repulsive Coulomb interaction. It is possible
then, that the appearance of the magic numbers for three-particle systems with repulsive interactions is
simply a manifestation of the combination of even/odd oscillations in the energy
with the pattern of degeneracy for three identical particles.

\section{Summary}

The three-boson problem in 2D in the presence of a transverse magnetic field
is suprisingly well described using the adiabatic hyperspherical method. The
full energy spectrum presents very narrow avoided crossings between the
adiabatic energies, and away from these crossings the couplings between
channels can be largely ignored to a good approximation. This indicates that
the system is nearly separable in the hyperspherical picture. The adiabatic
hyperspherical picture provides a useful interpretation of transitions in
which excitations between levels can be achieved through either a hyperangular
excitation in which the internal configuration of the three-boson system is
changed or through a hyperradial vibrational excitation in which the internal
structure of the system remains the same. The adiabatic hyperangular
eigenvalues, $\varepsilon_{nM}\left(  R\right)  $, are exactly the same as
those found for three interacting bosons in free-space. The inclusion of the
magnetic field results in the addition of an effective isotropic trap, and an
angular momentum dependent shift.

When interacting via the s-wave pseudopotential, three-body states transition
from the weakly repulsive regime ($a\ll l_{c}$) to the weakly attractive
regime ($a\gg l_{c}$) as a function of the 2D scattering length. States that
interact via the s-wave interaction display an even/odd parity oscillation as
a function of the total internal angular momentum $M$. For small scattering
length, this parity oscillation combined with the fact that there is no $M=1$
lowest Landau level means that the lowest interacting three-boson state has
total internal angular momentum $M=3$. At higher values of angular momentum
the lowest Landau level becomes degenerate and a set of non-interacting states
emerge in which the inter-particle angular momentum has no $m=0$ component.
Interestingly, if the same pattern of even/odd parity oscillations persits
when higher partial wave interactions are included, in combination with the
pattern of degeneracy for the lowest Landau level, this might be the source of
the \textquotedblleft magic number\textquotedblright\ behavior seen in
three-particle systems interacting via long-range Coulomb interactions, and is
the subject of ongoing work.



\end{document}